\documentstyle{article}
\begin{document}
\newcommand{\parsq}[1]{\frac{\partial^{2}}{\partial #1^{2}}}

\vskip-1.5cm
\hskip12cm{\large HZPP-9802}

%%%%%%%%%%%%%%%%%%%%%%%%% ? %%
\hskip12cm{\large Jan 25, 1998}
%%%%%%%%%%%%%%%%% ????? %%%%%%
\vskip1cm
 \date{}
 \centerline{\huge When to carry out analytic continuation?}
%%%  \centerline{\huge     $^{*}$}
\bigskip
\bigskip

\centerline{Jian Zuo}
\centerline{\small Institute of Particle Physics, Huazhong Normal University}
\centerline{\small Wuhan 430070, People's Republic of China}
\centerline{\small Email: zuoj@iopp.ml.org}
\vspace{0.3cm}
\centerline{Yuan-Xing Gui}
\centerline{\small Department of Physics, Dalian University of Technology}
\centerline{\small Dalian 116023, People's Republic of China}
\centerline{\small Email: guiyx@dlut.edu.cn}
\bigskip

\begin{center} \begin{minipage}{125mm} \vskip 0.5in
\begin{center}{\bf Abstract }\end{center}
{ This paper discusses the analytic continuation in the thermal
field theory by using the theory of $\eta-\xi$ spacetime. Taking a simple
model as example, the $2\times 2$ matrix real-time propagator is solved from
the equation obtained through continuation of the equation for the 
imaginary-time propagator. The geometry of the $\eta-\xi$ spacetime plays
important role in the discussion.}

\end{minipage}
\end{center}
\vskip1.5cm

%% {\bf PACS:} 25.75.-q
%% \vfill
%%  \noindent{* This work is supported in part by the NNSF of China}
\vskip1truecm

\newpage
\par It is well known that the finite temperature field theory (FTFT) has
two formalisms: the imaginary-time formalism and the real-time formalism 
(Landsman 1987). The imaginary-time formalism is charactered by the 
periodicity of imaginary-time which leads to discrete imaginary energy in the
imaginary-time thermal Green functions (Matsubara 1955, Fetter and Walecka 1965), and the 
real-time formalism is charactered by the doubling of the degrees of freedom
which causes the real-time thermal Green functions to have $2\times 2$ matrix
structures (Niemi and Semenoff 1984, Umezawa et al 1982). This paper discusses a problem concerning the
connection of the two formalisms, i.e. the analytic continuation of thermal
propagators. Although there is not direct analytic continuation between the
imaginary-time thermal propagators and the $2\times 2$ form
real-time thermal propagators even in the most simple case, such a relation
does exist between the equations for propagators. It is interesting to find
that the difference between these two situations is easily explained by the
geometrical features of a spacetime with $S^1$ topology, named the $\eta-\xi$
spacetime (Gui 1988, Gui 1990, Gui 1992, Gui 1993) rather than by the Minkowskian spacetime.
\par The theory of $\eta-\xi$ spacetime is constructed in order to provide a
unique geometrical background for FTFT. The most important parts of the
$\eta-\xi$ spacetime are its Euclidean section and Lorentzian section. The
Euclidean section has a $S^1$ topology, which makes quantum fields satisfy
the periodicity for imaginary-time, and the Euclidean propagators in the
$\eta-\xi$ spacetime correspond to the imaginary-time thermal propagators. An
interesting fact about the Lorentzian section is that its geometrical structure
is very much similar to those of the Rindler spacetime and black hole. The
infinities of the Minkowskian spacetime become "horizons" on the Lorentzian
section which lead to the doubling of degrees of freedom of fields, and the
vacuum propagator in the $\eta-\xi$ spacetime is equal to the $2\times 2$
matrix real-time thermal propagator in the Minkowskian spacetime. It was
suggested (Gui 1993, Zuo and Gui 1995) that the field theories on the Euclidean section
and Lorentzian section correspond to the imaginary-time formalism and
real-time formalism of FTFT, respectively. The special geometrical structures
of the $\eta-\xi$ spacetime shall also affect the relation between the two
formalisms of FTFT, e.g. the situation in which direct analytic continuation
can be carried out.
\par This paper first gives a brief description of the structures of the 
$\eta-\xi$ spacetime and some relations to be used. Then by discussing the
procedures of solving the equations for propagators on the Euclidean section
and on the Lorentzian section respectively, it explains how the geometry
influences the feasibility of analytic continuation.
\par The four dimensional $\eta-\xi$ spacetime can be regarded as the maximal
analytic complex extension of $S^{1}\times R^{3}$ manifold (Gui 1990). It
has the following complex metric:
 \begin{equation}                    
 ds^{2}=\frac{1}{\alpha^{2}(\xi^{2}-\eta^{2})}(-d\eta^{2}+d\xi^{2})+dy^{2}+dz^{2}
 \end{equation}
where $\alpha=2\pi\slash\beta$ is a real constant and $\eta, \xi, y, z$ are
complex variables. If we limit $\xi, y, z$ to be real and $\eta$ to be a pure
imaginary variable $i\sigma$, the Euclidean section of the $\eta-\xi$
spacetime is obtained:
 \begin{equation}
 ds^{2}=\alpha^{-2}(\xi^{2}+\sigma^{2})^{-1}(d\sigma^{2}+d\xi^{2})+dy^{2}+dz^{2}
 \end{equation}
which under the transformation
 \begin{eqnarray}
 \sigma &=& \alpha^{-1}e^{\alpha x}\sin\alpha \tau \nonumber\\
 \xi    &=& \alpha^{-1}e^{\alpha x}\cos\alpha \tau
 \end{eqnarray}
becomes a flat Euclidean spacetime
 \begin{equation}
 ds^{2} = d\tau^{2}+dx^{2}+dy^{2}+dz^{2}
 \end{equation}
The metric (2) is singular at $\sigma=\xi=0$, so it describes an Euclidean 
spacetime with $S^{1} \times R^{3}$ topology. The periodicity of polar angle
$\alpha \tau$ naturally supplies the periodicity of imaginary-time in FTFT.
Now continuate $\sigma$ to $\sigma e^{-i\theta}$. The singularity becomes
 \begin{equation}
 \xi^{2} + \sigma^{2} e^{-2i\theta}=0
 \end{equation}
which requires $\sigma=\xi=0$ for all values of $\theta$ except 0 and
$\pi \slash 2$. $\theta=0$ is just the case of the Euclidean section. When
$\theta=\pi \slash 2$, the $\sigma$ coordinate changes into a real $\eta$
coordinate and the resulted $\eta-\xi$ hyperplane is the Lorentzian
section of the $\eta-\xi$ spacetime. The singularities on the Lorentzian
section are described by
 \begin{equation}
 \xi^{2} - \eta^{2} =0
 \end{equation}
which divide the Lorentzian section into four disjointed regions I, II, III,
IV. This structure resembles that of the Schwarzschild spacetime, thus the
singularities (6) are also called "horizons". Each of the regions is
identified with a four dimensional Minkowskian spacetime. One can see this
from the transformation
 \begin{equation}
 \eta = \alpha^{-1} e^{\alpha x} \sinh \alpha t \hspace{2cm}
 \xi  = \alpha^{-1} e^{\alpha x} \cosh \alpha t
 \end{equation}
which transforms region I of the Lorentzian section into
 \begin{equation}
 ds^{2}=-dt^{2}+dx^{2}+dy^{2}+dz^{2}
 \end{equation}
Similarly, regions II, III, IV are transformed by
 \begin{eqnarray}
 \eta &=& -\alpha^{-1} e^{\alpha x} \sinh \alpha t \hspace{2cm}
 \xi  = -\alpha^{-1} e^{\alpha x} \cosh \alpha t \nonumber \\
 \eta &=&  \alpha^{-1} e^{\alpha x} \cosh \alpha t \hspace{2cm}
 \xi  =  \alpha^{-1} e^{\alpha x} \sinh \alpha t \nonumber \\
 \eta &=& -\alpha^{-1} e^{\alpha x} \cosh \alpha t \hspace{2cm}
 \xi  = -\alpha^{-1} e^{\alpha x} \sinh \alpha t
 \end{eqnarray}
respectively. The appearence of singularities (6) and the existence of several
regions make it possible for the Lorentzian section to explain the doubling of
degrees of freedom. While the original degrees of freedom are provided by
region I, the additional degrees of freedom can be supplied by region II.
\par There is a relation between the transformations of regions I and II:
 \begin{eqnarray}
 \eta &=& -\alpha^{-1} e^{\alpha x} \sinh{\alpha t} = \alpha^{-1} e^{\alpha x}
          \sinh{\alpha (t-i\beta \slash 2)} \nonumber\\
 \xi  &=& -\alpha^{-1} e^{\alpha x} \cosh{\alpha t} = \alpha^{-1} e^{\alpha x}
          \cosh{\alpha (t-i\beta \slash 2)}
 \end{eqnarray}
Using Minkowskian coordinates $(t,x,y,z)$, the relation (10) becomes the
relation between a point $a_1 (t_1, x_1, y_1, z_1)$ in region I and its 
reflected point $a_2 (t_2, x_2, y_2, z_2)$ in region II:
 \begin{equation}
 t_2 = t_1 -i\beta \slash 2 \hspace{1cm} x_2 = x_1 \hspace{1cm} 
 y_2 = y_1 \hspace{1cm} z_2 = z_1
 \end{equation}
i.e. their Minkowskian coordinates differ only in an imaginary-time interval
$i\beta\slash2$.
\par Another important relation is that the direction of time $t$ in region II
is against the time direction in region I. The time-like Killing fields on the
Lorentzian section are defined by (Gui 1990):
 \begin{equation}
 (\frac{\partial}{\partial \lambda})^a = \varepsilon \alpha (\xi \eta^{a} + \eta \xi^{a})
 \end{equation}
where $\varepsilon = 1$ and -1 for region I and II respectively. It is natural
to choose the Killing parameter $\lambda$ as the time coordinate, which coincide
with Minkowskian time coordinate $t$ in region I, and $-t$ in region II.
\par Using the above relations, one can see how the geometry of the $\eta-\xi$
spacetime gets into effect on the problem of analytic continuation. The
following discussion takes the example of the massless free scalar field in
two dimensional $\eta-\xi$ spacetime. Simple as it is, it makes one touch
directly the physical and geometrical essence and avoid difficult technical
details. The equation for propagators of this field on the Euclidean section is
 \begin{equation}
 (\frac{\partial^{2}}{\partial \sigma^{2}}+\frac{\partial^{2}}{\partial \xi^{2}})
 D_I (A-A') = -(-g_E)^{-1\slash 2}\delta(A-A')
 \end{equation}
where $g_E$ stands for the determinant of metric of the Euclidean section and 
$D_I$ is the imaginary-time propagator. Under transformation (3), the equation
becomes:
 \begin{equation}
 (\parsq{\tau} + \parsq{x}) D_I(A-A') = -\delta (A-A')
 \end{equation}
in which the points A and A' have the coordinates $(\tau,x)$ and $(\tau',x')$
respectively. Since the Euclidean section has $S^1$ topology, the propagators
naturally satisfy the periodicity boundary condition:
 \begin{equation}
 D_I (\tau -\tau') = D_I (\tau -\tau' +\beta)
 \end{equation}
which is just the KMS condition (Kubo 1957, Martin and Schwinger 1959). Using this condition,
the imaginary-time thermal propagator in momentum space is routinely obtained
(Fetter and Walecka 1965):
 \begin{equation}
 D_I(\omega_n, k) = \frac{1}{\omega_n^2 +k^2}
 \end{equation}
where $\omega_n = 2\pi n \slash -i\beta$.
\par Now continuate the equation (13) to the equation on the Lorentzian
section. This is done by continuate $\sigma$ to $i\eta$:
 \begin{equation}
 [-\parsq{\eta}+\parsq{\xi}]D_R(A-A') = -(-g_L)^{-1\slash2}\delta(A-A')
 \end{equation}
where $g_L$ stands for the determinant of metric of the Lorentzian section. It
is noticed that, since the whole Lorentzian section is obtained after a
continuation of the Euclidean section, the points A and A' can be located in 
every one of the four regions. However, the regions III and IV are spacelike
with respect to regions I and II, so one only need to consider the cases that
A and A' are located in regions I and II. Using the Minkowskian coordinates 
$(t, x)$, the equation (17) can be transformed into
 \begin{equation}
 [-\parsq{t}+\parsq{x}]D_R(A-A') = -\delta(A-A')
 \end{equation}
Since both of A and A' can be located in each of the two regions, there are
four cases. If one perform a Fourier transformation of equation (18), there
will be four expressions for the field modes. Using the suffix 1, 2 to stand
for coordinates in regions I and II respectively, one can write the expressions
of field modes as:
 \begin{eqnarray}
 e_{11} &=& \exp \lbrace ik(x_1-x_1')-ik_0(t_1-t_1')\rbrace \hspace{1cm}
 e_{12} = \exp \lbrace ik(x_1-x_2')-ik_0(t_1-t_2')\rbrace \nonumber \\
 e_{21} &=& \exp \lbrace ik(x_2-x_1')-ik_0(t_2-t_1')\rbrace \hspace{1cm}
 e_{22} = \exp \lbrace ik(x_2-x_2')-ik_0(t_2-t_2')\rbrace
 \end{eqnarray}
It shall be noted that $e_{12}$ and $e_{21}$ are only formally like plane waves.
But we can view the coordinates $t$ and $x$ in these two expressions as functions
of the coordinates $\eta$ and $\xi$, thus $e_{12}$ and $e_{21}$ represent the
field modes on the whole Lorentzian section which relate different regions.
\par The Fourier coefficients $D(k, k_0)$ shall be different for different
field modes. Hence the transformed equation takes a form of matrix:
 \begin{eqnarray}
 \int dk dk_0 (k^2-k_0^2)
  \Bigg(\begin{array}{cc}
   D_{11}e_{11} & D_{12}e_{12}\\
   D_{21}e_{21} & D_{22}e_{22}
  \end{array}\Bigg) 
 = -\int dk dk_0\Bigg(
                \begin{array}{cc}
                 e_{11} \hspace{0.5cm} e_{12}\\
                 e_{21} \hspace{0.5cm} e_{22}
                \end{array}
                \Bigg)
 \end{eqnarray}
By the use of coordinate relation
 \begin{equation}
 t_2 =t_1 -i\beta\slash2 \hspace{3cm} x_2 =x_1
 \end{equation}
one rewrites the matrix on the right side of (20) as
 \begin{eqnarray}
 \Bigg(
   \begin{array}{cc}
      e^{ik(x_1-x_1')-ik_0(t_1-t_1')} \hspace{1cm} e^{ik(x_1-x_1')-ik_0(t_1-t_1'+i\beta\slash2)}\\
      e^{ik(x_1-x_1')-ik_0(t_1-t_1'-i\beta\slash2)} \hspace{1cm} e^{ik(x_1-x_1')+ik_0(t_1-t_1')}
   \end{array}
 \Bigg)
 \end{eqnarray}
A sign is changed in 2-2 component because the direction of time in region I
points against the one in region II. 
\par The term $i\beta\slash2$ in the off-diagonal components is explained as
thermal factor caused by the geometrical structure of the Lorentzian section.
Since regions I and II are separated on the Lorentzian section by the
"horizons", they can only be connected by complex pathes which run along half
circles on the Euclidean section and result in the imaginary-time interval
$i\beta\slash2$. Physically, the field modes $e_{12}$ and $e_{21}$ can not be
completely measured by the observors in region I, to whom the information in
region II is screened by the "horizons". It is just this lost of information
that makes the observors in region I find a finite temperature. In this
explanation, the multiple-region geometrical structure, and thus the doubling
of degrees of freedom, are the origins of thermal factor. One notices that
this geometry-induced thermal effect is quite similar to the gravity-induced
Hawking-Unruh effects (Hawking 1974, Unruh 1975).
\par Using similar coordinate relation on the left side of (20), one gets
four equations:
 \begin{eqnarray}
 (-\parsq{t}+\parsq{x})D_{11}(t-t', x-x') &=& -\delta (t-t', x-x') \nonumber\\
 (-\parsq{t}+\parsq{x})D_{12}(t-t'+i\beta\slash2, x-x') &=& -\delta (t-t'+i\beta\slash2, x-x') \nonumber\\
 (-\parsq{t}+\parsq{x})D_{21}(t-t'-i\beta\slash2, x-x') &=& -\delta (t-t'-i\beta\slash2, x-x') \nonumber\\
 (-\parsq{t}+\parsq{x})D_{22}(t'-t, x-x') &=& -\delta (t'-t, x-x') 
 \end{eqnarray}
Here the suffixes 1 for coordinates are omitted. By solving these equations 
one get
 \begin{eqnarray}
 D_{\beta}(k_0, k)=\Bigg(
                   \begin{array}{cc}
                    D_{11} \hspace{1cm} D_{12}\\
                    D_{21} \hspace{1cm} D_{22}
                   \end{array}
                   \Bigg)
 \end{eqnarray}
where
 \begin{equation}
 D_{11}=\frac{1}{k_0^2-k^2+i\epsilon}-\frac{2\pi i\delta(k_0^2-k^2)}{e^{\beta k_0}-1}
 \end{equation}
 \begin{equation}
 D_{12}=\frac{2\pi ie^{-\beta k_0\slash2}\delta(k_0^2-k^2)}{1-e^{-\beta k_0}}
 \end{equation}
and 
 \begin{equation}
 D_{21}=D_{12} \hspace{3cm} D_{22}=-D^*_{11}
 \end{equation}
Thus the solution of the equation (17), which is obtained by continuation of
the equation for the imaginary-time thermal propagator, is just the $2\times
2$ matrix real-time thermal propagator.
\par It may seem contradictory that propagators with so diverse forms that
they can not be directly continuated to each other are solutions of equations
which can be mutually obtained by continuation. This can be explained by 
reviewing the roles of different geometrical structures of the Euclidean and
Lorentzian sections in the processions of solving the equations.
\par Since each of the equations (13) and (17) holds on the whole Euclidean
section or Lorentzian section respectively, and no additional singularity is
met while the sections rotate to each other, the analytic continuation between
the equations is feasible. But direct analytic continuation between the
thermal propagators is hindered by the different singularities on the two
sections, which impose different requirements in the courses of solving these
equations. On one hand, the singularity $\sigma = \xi =0$ results in the 
$S^1$ topology of the Euclidean section and thus the periodicity boundary
condition (15), which requires the propagators on this section to be 
transformed into Fourier serie. On the other hand, the Lorentzian section is
divided into four disjointed regions by the "horizons" (6), which cause
different expressions of field modes and thus give four matrix components of
the thermal propagator.
\par One remembers an early work (Dolan and Jackiw 1974) which tried to get the real-time
thermal propagators through direct analytic continuation of the imaginary-time
thermal propagators. However, the result was only the 1-1 component of the
$2\times 2$ real-time thermal propagator, which leads to difficulties in
calculations. 
\par In view of the analysis in this paper, it is also clear to see why the
early continuation was not complete. In fact, there was an implicit premise
for that attempt. It was supposed that the background spacetime for the
real-time formalism is the Minkowskian spacetime. This corresponds to only a
region of the Lorentzian section of the $\eta-\xi$ spacetime, hence it can
not provide the doubling of degrees of freedom. The 1-1 component of the
propagator is just the case that both A and A' are located in region I, while
the other cases are ignored in that attempt. Since the singularities on the
Lorentzian section, which play important role in solving the equation for the
real-time propagator, had not been considered, it is not strange that the
$2\times2$ matrix real-time thermal propagator could not be obtained.
\par It shall be noticed that this paper takes as example a very simple model,
and its extension to more practical calculation is far from straightforward.
However, by discussing the influence of spacetime geometry on the feasibility
of analytic continuation, it suggests a new intuitive way of looking at this
problem.
\par \mbox{}\\
\bf{Acknowledgement}\\
This work is supported by the National Natural Science Foundation of China.\\
\mbox{}\\
Dolan L., Jackiw R., Phys. Rev. D9(1974)3320.\\
Fetter A. L., Walecka J. D., Quantum Theory of Many-Particle System, 
                             McGraw-Hill Book Company, 1965.\\
Gui Y. X., Sci. Sin. 31A(1988)1104.\\
Gui Y. X., Phys. Rev. D42(1990)1988.\\
Gui Y. X., Phys. Rev. D46(1992)1869.\\
Gui Y. X., Sci. Sin. 36A(1993)571.\\
Hawking S. W., Commun. Math. Phys. 43(1975)199.\\
Kubo R., J. Phys. Soc. Japan 52(1957)570.\\
Landsman N. P., van Weert Ch. G., Phys. Rep. 145(1987)141.\\
Martin P. C., Schwinger J., Phys. Rev. 115(1959)1342.\\
Matsubara T., Prog. Theor. Phys. 14(1955)351.\\
Niemi A. J., Semenoff G. W., Ann. Phys. 152(1984)105.\\
Umezawa H., Matsumoto H., Tachiki M., Thermo Field Dynamics and Condensed
                                      States, North-Holland, 1982.\\
Unruh W. G., Phys. Rev. D14(1976)870.\\
Zuo J., Gui Y. X., J. Phys. A: Math. Gen. 28(1995)4907.

\end{document}